\def \SAIT #1 #2 {{\em Mem.\ Soc.\ Astron.\ It.\/} {\bf #1}, #2}
\def \MESS #1 #2 {{\em The Messenger\/} {\bf #1}, #2}
\def \ASTRNACH #1 #2 {{\em Astron. Nach.\/} {\bf #1}, #2}
\def \AAP #1 #2 {{\em Astron. Astrophys.\/} {\bf #1}, #2}
\def \AAL #1 #2 {{\em Astron. Astrophys. Lett.\/} {\bf #1}, L#2}
\def \AAR #1 #2 {{\em Astron. Astrophys. Rev.\/} {\bf #1}, #2}
\def \AAS #1 #2 {{\em Astron. Astrophys. Suppl. Ser.\/} {\bf #1}, #2}
\def \AJ #1 #2 {{\em Astron. J.\/} {\bf #1}, #2}
\def \ANNREV #1 #2 {{\em Ann. Rev. Astron. Astrophys.\/} {\bf #1}, #2}
\def \APJ #1 #2 {{\em Astrophys. J.\/} {\bf #1}, #2}
\def \APJL #1 #2 {{\em Astrophys. J. Lett.\/} {\bf #1}, L#2}
\def \APJS #1 #2 {{\em Astrophys. J. Suppl.\/} {\bf #1}, #2}
\def \APSS #1 #2 {{\em Astrophys. Space Sci.\/} {\bf #1}, #2}
\def \ASR #1 #2 {{\em Adv. Space Res.\/} {\bf #1}, #2}
\def \BAIC #1 #2 {{\em Bull. Astron. Inst. Czechosl.\/} {\bf #1}, #2}
\def \JSQRT #1 #2 {{\em J. Quant. Spectrosc. Radiat. Transfer\/} {\bf #1}, #2}
\def \MN #1 #2 {{\em Mon. Not. R. Astr. Soc.\/} {\bf #1}, #2}
\def \MEM #1 #2 {{\em Mem. R. Astr. Soc.\/} {\bf #1}, #2}
\def \PLR #1 #2 {{\em Phys. Lett. Rev.\/} {\bf #1}, #2}
\def \PASJ #1 #2 {{\em Publ. Astron. Soc. Japan\/} {\bf #1}, #2}
\def \PASP #1 #2 {{\em Publ. Astr. Soc. Pacific\/} {\bf #1}, #2}
\def \NAT #1 #2 {{\em Nature\/} {\bf #1}, #2}

\documentstyle[psfig]{memsait}
\input epsf.sty
\begin{opening}
\title{NUCLEOSYNTHESIS IN INTERMEDIATE MASS AGB STARS} 
\author{John Lattanzio$^{1,2}$, Manuel Forestini$^2$, Corinne Charbonnel$^3$}
\institute{$^1$Department of Mathematics and Statistics, Monash University, Clayton, Victoria, 3168, Australia\\
$^2$Laboratoire d'Astrophysique, Universit\'e Joseph Fourier, BP 53, 
F-38041 Grenoble Cedex 9, France\\
$^3$Laboratoire d'Astrophysique de l'Observatoire Midi-Pyr\'en\'ees,
CNRS UMR 5572, 
14 Av E. Belin, 31400 Toulouse, France}
\date{} 
\end{opening}

\begin{document}

\oddpagefooter{}{}{} 
\evenpagefooter{}{}{} 
\ 
\bigskip

\begin{abstract}
We present a summary of the main sites for nucleosynthesis in intermediate
mass Asymptotic Giant Branch (AGB) stars. We then discuss some detailed 
evolutionary models and how these have been used to create a synthetic evolution 
code which calculates the nucleosynthesis very rapidly, enabling us to 
investigate changes in some uncertain parameters in AGB evolution, such
as mass-loss and dredge-up. We then present results for C, C/O, Mg and Al.
We also discuss the changes due to the recent NACRE compilation of reaction rates.
\end{abstract}

\section{Introduction}
Asymptotic Giant Branch (AGB) stars show a rich variety of nucleosynthesis
processes. There are two primary sites for these processes: the helium burning
shell and the hydrogen burning shell. 
The reason that these are of such
interest in an AGB star is that the helium shell shows periodic
thermal excursions, and the hydrogen shell overlaps the bottom of
the convective envelope (in some cases), producing what is called
``hot bottom burning'' (HBB). 

\subsection{The Helium Shell}
The main nucleosynthesis occurring in the helium
shell is the production of alpha elements (via helium burning) as
well as neutron capture reactions. The details of these processes depend
primarily on the mass of the hydrogen exhausted core, hereafter $M_H$,
and the so-called $^{13}$C-pocket which is believed to form as a result of
partial mixing of the convective envelope with carbon-enriched matter,
during the dredge-up phase (Iben \& Renzini 1982a,b; Herwig 1999). 
As many of the star's characteristics
are determined primarily by $M_H$ it is possible to parametrise the detailed
models in terms of $M_H$ (and other variables) and then run ``synthetic''
evolutionary models. 
This allows us to explore the domains of uncertainties of the input physics
of the models. Indeed, 
detailed models are extremely time consuming,
and to the extent that these parametrisations represent the main
features of the evolution, they are very useful in determining the
behaviour of populations of AGB stars, which would not be possible
if we were forced to use only full evolutionary and nucleosynthesis 
calculations.


\subsection{The Hydrogen Shell: Hot Bottom Burning}
In stars more massive than about $5M_\odot$ the bottom of the convective
envelope dips into the top of the hydrogen burning shell. This then
brings a fresh supply of nuclides into the burning region, and mixes the
products of these reactions to the surface. The temperature at the
bottom of the envelope reaches as high as 100 million degrees, and
hence not only the CNO cycles but also the Ne-Na and Mg-Al chains/cycles
are substantially active. These produce changes in the photospheric
composition which are visible via spectroscopy. Unfortunately, the
temperature at the base of the convective envelope $T_{bce}$ is dependent 
not only on the core mass $M_H$ but also on the envelope mass $M_{env}$ as 
well as the stellar composition. Such stars are not as easily modeled by 
synthetic calculations.

\section{Detailed Models}
Recently Frost (1997) has calculated the detailed evolution of
nine cases, for masses of $4$, $5$ and $6M_\odot$ each with metal content
$Z=0.02$, $0.008$ and $0.004$, appropriate to the solar neighbourhood,
and the Large and Small Magellanic Clouds, respectively. These evolutionary
calculations began before the main sequence and went through to the
end of the AGB, as was determined by a failure of the evolution code
to converge on further models. This appears to be due to the reduction
of gas pressure (and concomitant super-Eddington luminosity) as previously
found by Faulkner \& Wood (1985) and Sweigart (1998). Typically
there was about $1M_\odot$ of envelope still remaining at
this stage, which would presumably be ejected as a planetary nebula.

Each of these cases has been the subject of a detailed nucleosynthesis
study using a heavily modified post-processing code (Cannon 1993)
which include 74 species and over 500 reactions. 
Some results from these models have been published (eg Frost et al 1998,
Lattanzio \& Forestini 1998) but
this paper presents the first of a series examining the detailed
results.

\section{Synthetic Models}
Our present approach is to use the detailed stellar models of Frost (1997) as
the basis of synthetic models, using the synthetic
code of Forestini \& Charbonnel 
(1997). In this way we force the synthetic models to follow the
results of the detailed models (accuracy is within 10\%) yet we are
then free to vary those parameters which will not have substantial
feedback on the evolution, and determine (rapidly) the effect of such
changes. For example, changes in nuclear reaction rates are not expected
to have any effect on the evolution (unless they are for the major energy
producing reactions, such as the pp or CNO cycles) but they can produce 
substantial changes in the predicted envelope composition. These can be
checked very easily with the synthetic code. Likewise, changes
in the mass loss rates and the dredge-up law, if not too substantial, will
have a minimal effect on the evolution. 

The models presented here are the standard case, as considered by 
Frost (1997), but we also investigate the effect of the new NACRE 
compilation of reaction rates (Angulo et al, 1999).

\section{Standard Case}
We will present the results in the form of surface abundances, or isotopic
ratios, during the AGB lifetime of the nine cases discussed above. 
Each figure shows nine cases, on a 3$\times$3 grid with mass varying along
the $x$-axis and $Z$ decreasing along the $Y$-axis: in this way we expect
HBB to increase for increasing $x$ (mass) or increasing $y$ (decreasing $Z$).
The most extreme HBB is for the top right hand graph, the $M=6, Z=0.004$ case.

\begin{figure}
\centerline{\psfig{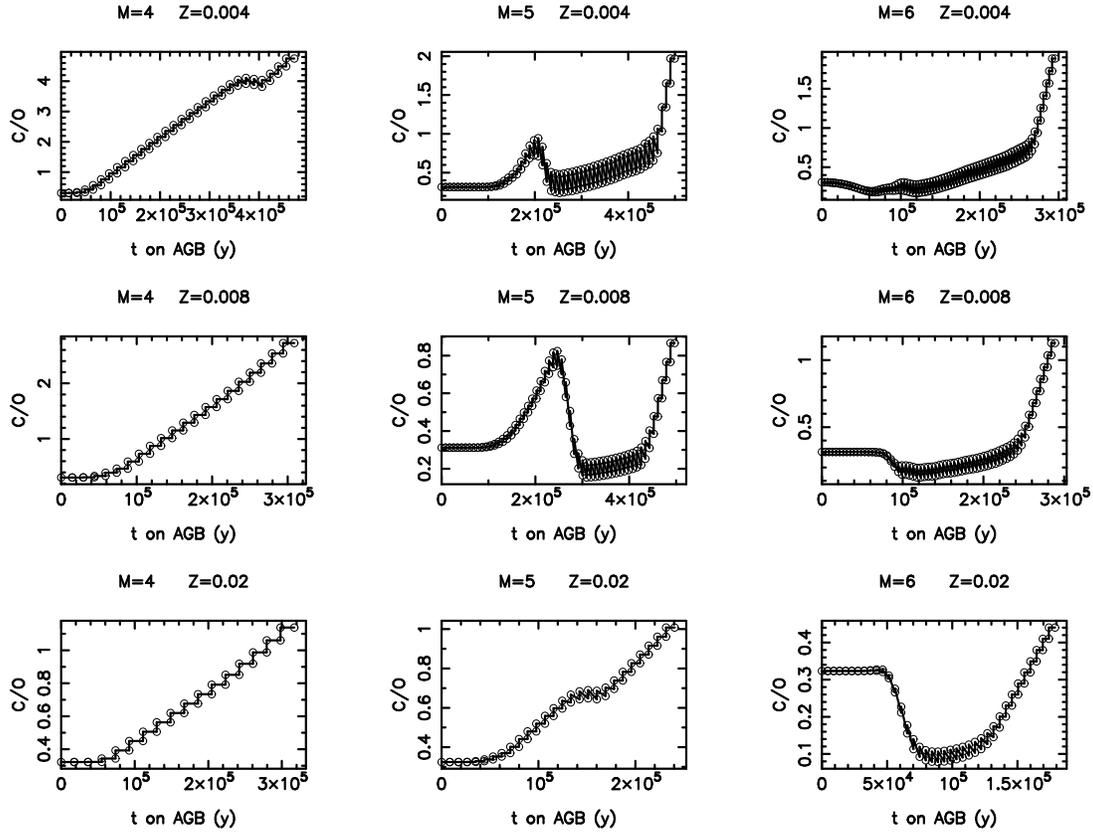}}
\caption[h]{Surface C/O ratios for the nine cases described in the text.}
\end{figure}

\subsection{Li}
We do not show here the results for $^7$Li, but note that they agree qualitatively
and quantitatively with those of Sackmann \& Boothroyd  (1992) as well
as Forestini \& Charbonnel (1997).
We also note that
the $M=4, Z=0.004$ case is both a Li-rich star 
(with $\epsilon(^7$Li$) \simeq 4$) and a Carbon star, 
for about $10^5$ years: for just under  half of this period (about $40,000$
years) it shows $^{12}$C/$^{13}$C$ < 10$ and would be classed as a J~star, but
earlier in the evolution this isotopic ratio was as high as 100.

\subsection{C/O}
Figure~1  shows the C/O ratio for the nine cases under consideration. Note
that dredge-up increases this ratio (by adding carbon to the stellar envelope)
and then HBB, where it exists, decreases it again as it transforms C into
N. We also see in this figure that HBB can prevent the formation of
C stars, in some cases, and in others it merely delays their
formation to a time when the HBB has stopped and the dredge-up
continues (see Frost et al 1998 for details).

\begin{figure}
\centerline{\psfig{figure=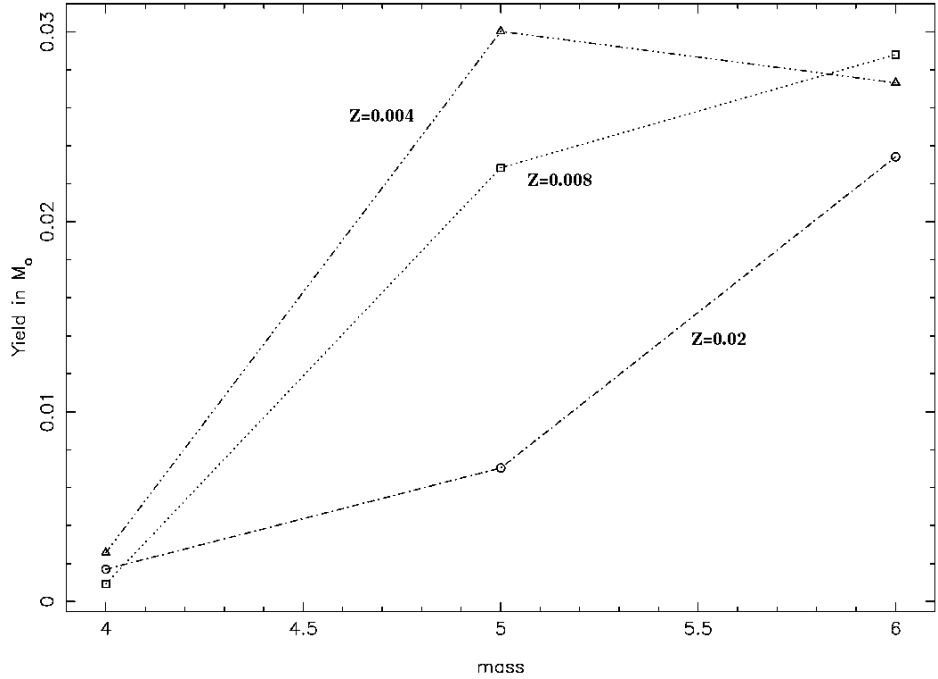,height=11cm}}
\caption[h]{Yields of $^{14}$N, in solar masses, for each case considered.}
\end{figure}

\subsection{$^{12}$C/$^{13}$C and $^{14}$N}
A useful indicator of HBB is the ratio of $^{12}$C/$^{13}$C, 
which increases due to
dredge-up but then decreases once HBB begins processing the envelope via
the CN cycle. So much material is cycled through the (thin) burning
region that in some cases
this carbon isotopic ratio reaches its equilibrium value of about
3.5. A consequence of this significant CNO cycling is that there is a substantial
amount of primary $^{14}$N produced, as shown in Figure~2, which plots the yield of $^{14}$N
for each of the nine cases. Note that up to $0.03M_\odot$ of $^{14}$N
is returned {\it per star\/} in extreme cases.

\begin{figure}
\centerline{\psfig{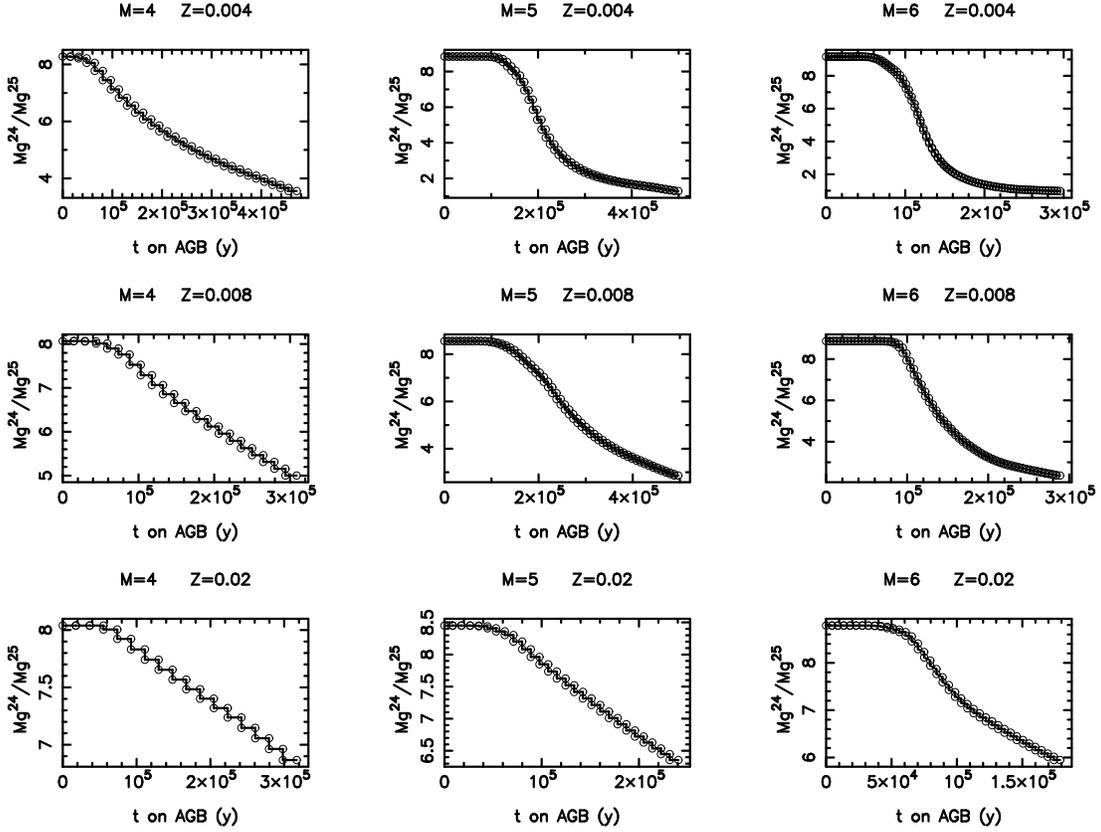}}
\caption[h]{Surface $^{24}$Mg/$^{25}$Mg ratios for the nine 
cases described in the text.}
\end{figure}

\subsection{Mg isotopes}
Of particular interest to us here is the production of $^{25}$Mg and $^{26}$Mg.
In the intermediate mass stars considered in this work, the main
neutron source in the helium shell is $^{22}$Ne$(\alpha,$n$)^{25}$Mg where 
most of the $^{22}$Ne
is made from successive alpha captures on $^{14}$N. Hence this 
$^{22}$Ne is primary,
as is much of the resulting $^{25}$Mg. 
Further, substantial $^{26}$Mg is also produced
by the Mg-Al cycle. Throughout, there is a negligible change in
the $^{24}$Mg abundance of the stars, so that overall the ratios of 
$^{24}$Mg/$^{25}$Mg
and $^{24}$Mg/$^{26}$Mg decrease from solar values of about 8 to about 1 or 2,
as shown in Figures~3 and 4. Note that this is mostly
primary $^{25}$Mg and $^{26}$Mg
that is produced.

\begin{figure}
\centerline{\psfig{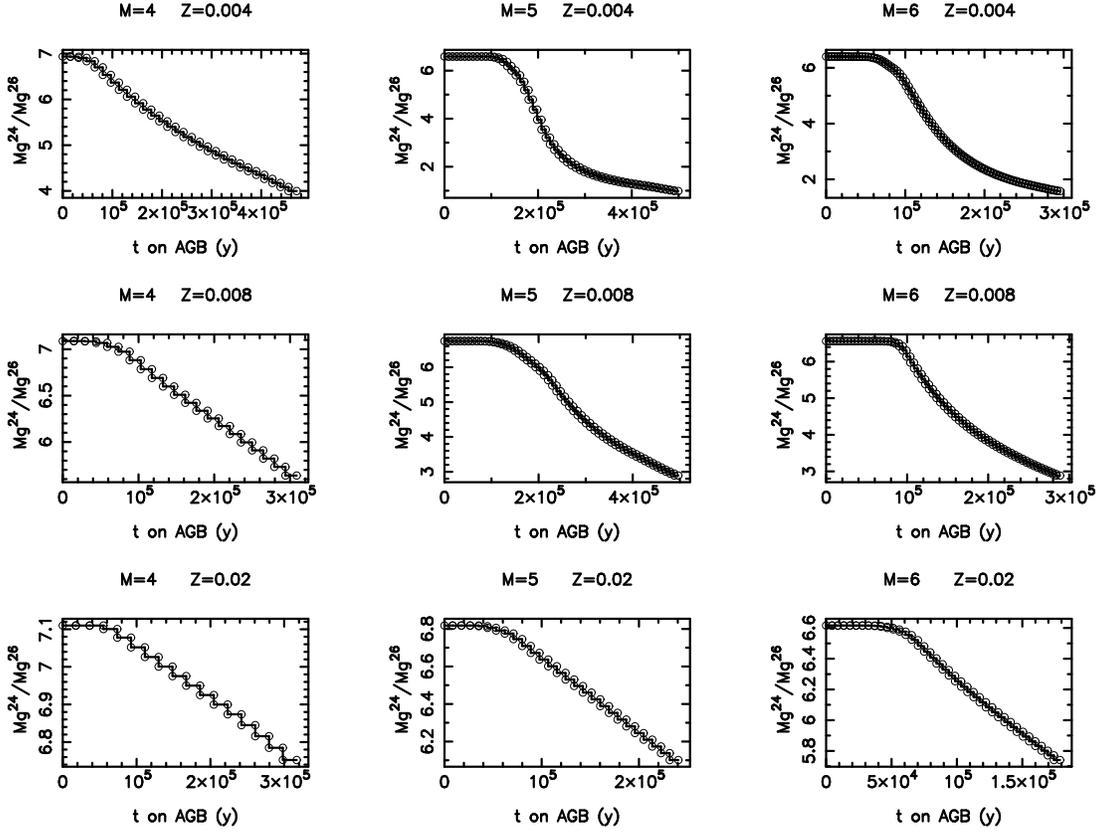}}
\caption[h]{Surface $^{24}$Mg/$^{26}$Mg ratios for the nine cases 
described in the text.}
\end{figure}

This is important for explaining the observed Al enhancements seen in
giants in globular clusters: to obtain the Al abundances as observed
it appears that substantial enhancements of   $^{25}$Mg and $^{26}$Mg   
are required (Denissenkov et al. 1998)
and the models presented here indicate that intermediate mass AGB stars
of low metallicity may produce such abundances. This will be the subject 
of further work.

\begin{figure}
\centerline{\psfig{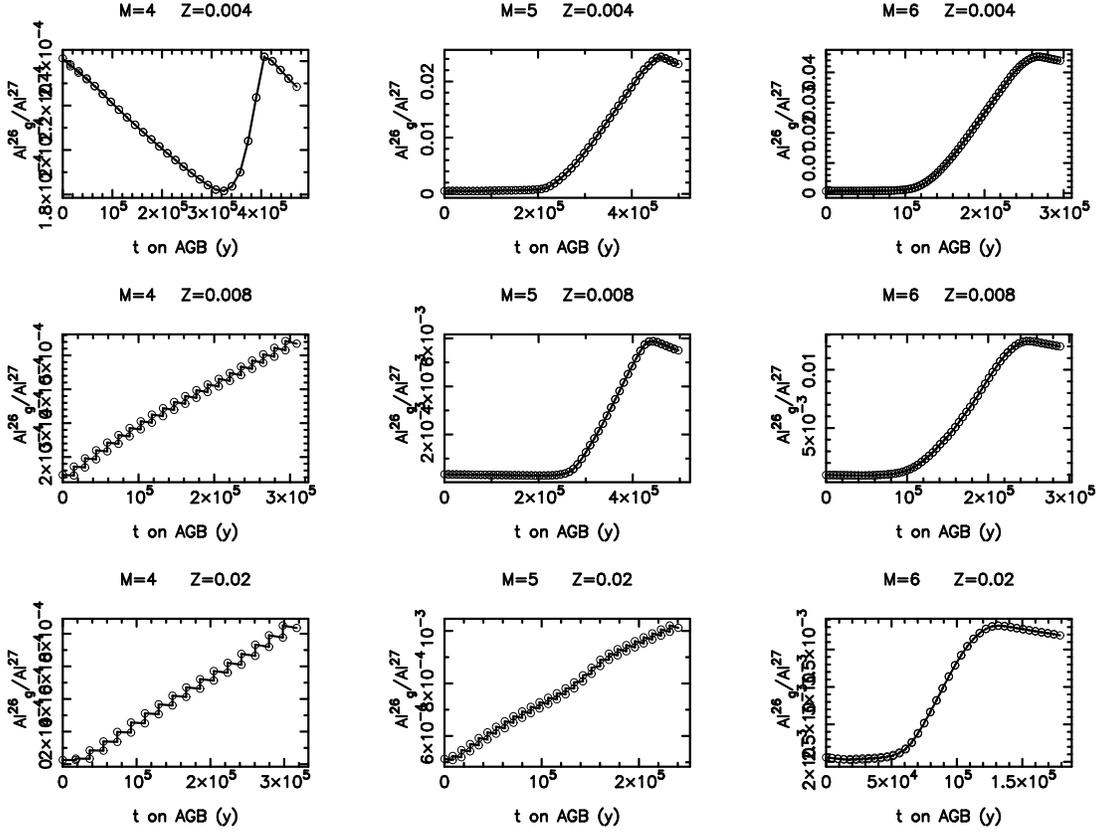}}
\caption[h]{Surface $^{26}$Al/$^{27}$Al ratios for the nine cases 
described in the text.}
\end{figure}

\subsection{$^{26}$Al/$^{27}$Al}
We show in Figure~5 the isotopic ratios of $^{26}$Al/$^{27}$Al in our stellar models.
The substantial HBB is responsible for producing a large amount of $^{26}$Al
and ratios of $^{26}$Al/$^{27}$Al as high as 
0.15 are obtained. This is in quantitative
agreement with the most extreme meteoritic grains which show ratios
as high as 0.10.

\section{NACRE rates}
The calculations presented so far use the same rates given in 
Forestini \& Charbonnel (1997). 
We have repeated these calculations with the new nuclear
reaction rates published by the NACRE consortium (Angulo et al
1999). The only
substantial change is a reduction in the amount of $^{26}$Al production, with 
a subsequent reduction on the $^{26}$Al/$^{27}$Al ratios of a factor of 2 to 4.
If this rate is correct then AGB stars are unlikely to be the source of the
most extreme $^{26}$Al/$^{27}$Al ratios seen in meteoritic grains. Such grains
may then be produced in WR stars. The rate needs further checking,
of course.

\section{Conclusions}
Intermediate mass AGB stars are important sources of primary $^{14}$N,
and produce Mg isotopic ratios that are far from solar. This could
be important for Al production in globular cluster red giants.


\acknowledgements
JCL thanks the Laboratoire d'Astrophysique de Grenoble
for their support and for their
hospitality, as well as the Universit\'e Joseph Fourier for 
financial assistance. He also thanks O'Callaghan's, Robert Monaghan and 
Lionel Siess for assistance through a difficult period! This work was 
funded (in part) by the Australian Research Council.



\end{document}